\documentclass[twocolumn, appendixfloats, revtex4, numberedappendix, twocolappendix, iop, letterpaper]{openjournal}
\usepackage[pass]{geometry}

\usepackage{natbib}

\usepackage[breaklinks,colorlinks,linkcolor=blue,citecolor=blue,urlcolor=blue]{hyperref}

\usepackage[T1]{fontenc}

\usepackage{orcidlink}

\usepackage{xspace}

\usepackage{graphicx}

\usepackage{paralist}
\newcommand{\HI}{\ifmmode \mathrm{H}\,\textsc{i} \else H~{\sc i}\fi\xspace}
\newcommand{\nhi}{\ifmmode \mathrm{N(H}\,\textsc{i}) \else N(H~{\sc i})\fi\xspace}
\newcommand{\mgii}{\ifmmode \mathrm{Mg}\,\textsc{ii} \else Mg~{\sc ii}\fi\xspace}
\newcommand{\civ}{\ifmmode \mathrm{C}\,\textsc{iv} \else C~{\sc iv}\fi\xspace}
\newcommand{\ciii}{\ifmmode \mathrm{C}\,\textsc{iii}] \else C~{\sc iii}]\fi\xspace}
\newcommand{\mgal}{\ifmmode M_* \else $M_*$\fi\xspace}
\newcommand{\modot}{\ifmmode M_\odot \else $M_\odot$\fi\xspace}

\shorttitle{Cool Gas in the CGM of Post Starburst Galaxies}
\shortauthors{Z. Harvey et al.}

\begin{document}

\title{\vspace{-0.8cm}Cool Gas in the Circumgalactic Medium of massive Post Starburst Galaxies\vspace{-1.5cm}}

\author{Zoe Harvey\,\orcidlink{0009-0009-9830-550X}$^{1}$,
  Sahyadri Krishna\,\orcidlink{0009-0000-5833-4116}$^{1}$,
  Vivienne Wild\,\orcidlink{0000-0002-8956-7024}$^{1}$,
  Rita Tojeiro\,\orcidlink{0000-0001-5191-2286}$^{1}$,
  Paul Hewett\,\orcidlink{0000-0000-0000-0000}$^{2}$\\
}

\affiliation{
  $^{1}$School of Physics and Astronomy, University of St Andrews, St Andrews, Fife KY16 8AZ, UK\\
  $^{2}$Institute of Astronomy, University of Cambridge, Madingley Road, Cambridge, CB3 0HA, UK\\}
\thanks{$^*$E-mail: \href{zh52@st-andrews.ac.uk}{zh52@st-andrews.ac.uk}}

% Abstract of the paper
\begin{abstract}
  Observing the interplay between galaxies and their gaseous surroundings is crucial for understanding how galaxies form and evolve, including the roles of long-lived cool gas reservoirs, starburst and AGN driven outflows.  We use stacked \mgii\ absorption lines in the spectra of background quasars to study the cool gas out to 9\,Mpc from massive quiescent, star-forming and post-starburst galaxies with stellar masses $\log_{10}(\mgal/\modot)\gtrsim11$ and $0.4 \lesssim z \lesssim 0.8$ selected from the Baryon Oscillation Spectroscopic Survey (BOSS) CMASS galaxies. Consistent with previous studies, we observe a decline in absorption strength indicating a decrease in cool gas content with increasing distance from the galaxies, as well as decreasing star formation rate of the galaxies. Beyond 1\,Mpc, this decline levels off to the same absorption strength in all galaxy types, suggesting a transition from the circumgalactic medium (CGM) to the intergalactic medium (IGM) at approximately the virial radius of the host dark matter haloes. We find that post-starburst galaxies, that have experienced a recent burst of star formation that has rapidly quenched, exhibit significantly stronger \mgii absorption within at least $\sim$300\,kpc than star-forming or quiescent galaxies of the same stellar mass.  Because post-starburst galaxies are a potentially significant pathway for the formation of quiescent elliptical galaxies, our results have wide reaching implications for understanding the mechanisms involved in quenching star formation in galaxies. We speculate that the excess cool gas absorption around post-starburst galaxies is related to their observed high velocity ($\sim$1000\,km/s) cool gas outflows. Thus, strong, short-lived bursts of star formation impact the CGM around galaxies on 100's of kpc distances and Gyr timescales.

\end{abstract}

\keywords{galaxies: intergalactic medium, starburst, star formation; quasars: absorption lines}

\maketitle %OJA

%%%%%%%%%%%%%%%%%%%%%%%%%%%%%%%%%%%%%%%%%%%%%%%%%%

%%%%%%%%%%%%%%%%% BODY OF PAPER %%%%%%%%%%%%%%%%%%

\section{Introduction}

% galaxy bimodality

Understanding the physical processes which cause the observed properties of the galaxy population around us is a key goal of modern astrophysics. One major set of questions revolves around how and why galaxies switch off their star formation, which they have been doing en-masse since ``cosmic noon'' at $1<z<3$ when the total star formation rate density of the Universe peaked \citep{Madau2014}. In search of answers, increasing amounts of effort has been put into understanding galaxies as part of the larger ecosystem around them, and in particular the characteristics of the reservoirs of cool ($10^4$K) gas that surround them \citep[see e.g.][for a review]{Tumlinson2017}. 

% intro to CGM

Atomic hydrogen detected via \HI 21\,cm line emission extends to several times the radius of stellar disks, providing information about gas in the immediate vicinity of the galaxy. However, observing this external cool gas reservoir is challenging. \HI emission is faint and until the Square Kilometre Array (SKA) starts operations in the next decade, resolved HI mapping is only feasible in significant numbers for massive HI rich disks in the local Universe \citep[e.g.][]{Maddox2021,Dutta2022}.  Truncated HI disks indicate a disruption of gas supply, for example as galaxies fall into clusters \citep[e.g.][]{ReynoldsWALLABY2022}, but building the statistics needed to study the causes of star formation cessation in the galaxy population more generally is challenging. 

% QSOALs

An alternative probe which can provide information on the cool gas reservoir out to the host halo’s virial radius and beyond are absorption lines detected in the spectrum of a bright background source, typically a QSO. These absorption lines are the shadow of the gas around the galaxy and, although often challenging to interpret, provide the best information we have on the extent and state of that gas. The association between QSO absorption lines (QSOALs) and galaxies has been studied for over 30 years \citep[e.g.][]{BergeronBoisse1991,Bergeron1992,steidel1995}, initially providing a hugely valuable method to identify candidate high redshift galaxies for follow up spectroscopy. 

% MgII 

While a wide variety of atomic transitions result in observable absorption features, some are more commonly studied than others. Singly ionised magnesium (\mgii) is a particularly well studied QSOAL, in part due to its strength and existence as a doublet in moderate resolution spectra, allowing easy identification without the need for additional lines. Its rest frame wavelength of $\lambda\lambda2796,\,2803$\AA\ also aligns perfectly with the huge optical spectroscopic QSO survey datasets collected over recent decades, allowing the identification of cool gas clouds around galaxies using a single diagnostic over a large fraction of cosmic history ($0.4\lesssim z \lesssim 1.8$).

With an ionisation potential of 15\,eV \citep{Morton2003}, close to that of \HI (13.6\,eV), \mgii traces the $\sim10^4$K (``cool''), dense and neutral gas of the CGM. Population number density arguments (i.e. the number of absorbers per unit redshift, $dN/dz$), directly observed associations and photoionisation modelling show that \mgii absorption line systems with rest-frame equivalent widths of the strongest line in the doublet $W_{\lambda2796}>0.3$\AA\ are typically associated with Hydrogen Lyman limit systems (LLS), i.e. are optically thick to hydrogen ionising photons, with $\log \nhi > 17.3\,{\rm cm}^{-2}$ \citep[e.g.][]{BergeronStasinska1986,Churchill2000a}. Weaker \mgii absorbers are far more numerous \citep{Churchill1999a}, and expected to trace optically thin gas clouds to lower column densities of $\log N(HI)\sim16\,{\rm cm}^{-2}$ \citep{Rigby2002}. 

% MgII interpretation

The most easily measurable quantity, the line strength (``equivalent width'', W) from moderate spectral resolution observations, tells us relatively little about the amount of \mgii gas causing the absorption as the lines are saturated. High spectral resolution observations show that strong absorption lines ($>0.3$\AA) are composed of multiple sub-components at slightly different velocities, and that the total equivalent width increases linearly with the number of sub-components \citep{PetitjeanBergeron1990,Churchill2000b}. Cloud temperatures for \mgii absorbers can be estimated from line widths in high resolution spectroscopy to be $\sim25,000$K \citep{Churchill1999a}.
Even with high spectral resolution, the size of the \mgii absorbing clouds is difficult to constrain from observations. Photoionisation equilibrium modelling provides some constraints on the volume density, particularly for clouds where absorption from Hydrogen and other metal line transitions can help constrain metallicity and the UV ionising radiation field strength. This in turn can constrain cloud sizes to be $\lesssim$100\,pc \citep[e.g.][]{BergeronStasinska1986, Rigby2002, Prochaska2009, Crighton2015, Lan2017}.  The coherence of absorbers around galaxies with multiple sightlines suggests that, whether the clouds are smooth and extended, or dense and clumpy, they are self similar on kpc scales \citep{Rubin2018}.

% MgII simulations

Since single, randomly distributed sight lines constitute the majority of observational data on the CGM and are intrinsically difficult to interpret in isolation, simulations are essential for contextualising \mgii absorption features and constraining the physical properties of the associated gas structures. For example, in massive ($\log(\mgal/\modot)>11$) galaxies in the cosmological Illustris TNG50 simulation \citep{Pillepich2019,Nelson2019}, \mgii absorbers arise from small ($\sim1$\,kpc diameter), dense ($-2.5 \lesssim \log n/{\rm cm}^{-3} \lesssim -1$ ) cool ($10^4$K) clouds within the virial radius of the galaxies, with a low volume filling factor, but high covering factor \citep{Nelson2020}. In these simulations, they are caused by strong density perturbations leading to thermal instability and cool clouds condensing from the hot medium, with the perturbations initially seeded by external effects such as infalling satellites. Individual cool clouds exchange material constantly with the warmer medium around them, and are typically infalling onto the central galaxy. However, the existence of low ionisation material, such as \mgii, is highly sensitive to photoionisation, either from the background radiation field or the presence of an active galactic nucleus \citep[e.g.][]{Liang2018,Hani2018,Appleby2021}, meaning many simulations do not observe cool CGM beyond the immediate vicinity of the galaxies. Simulations remain unable to capture both the true multiphase nature of the CGM and the cosmological context in which galaxies sit, and many disagree at a very fundamental level, such as the mechanism for seeding of clouds and how they survive the constant mixing with the warmer medium in which they sit \citep[see][for a review]{FaucherGiguere2023}. 

% MgII observations - individual 

Further observational constraints on the physical nature of the CGM can come from looking at the QSOALs together with the physical properties of their host galaxies. Initial follow-up observations of fields around \mgii absorbers identified normal star-forming galaxies within a projected distance of $\sim$60\,kpc  \citep[e.g.][]{BergeronBoisse1991,Steidel1994}. Multi-object, integral field and grism spectroscopy, as well as blue-sensitive spectrographs, have made positive identification of host galaxies easier, with catalogues now containing 100's-1000's of \mgii absorber-galaxy pairs with projected separations ranging from $\sim10-200$\,kpc \citep[e.g.][]{Nielsen2013, Huang2021, Lundgren2021, Dutta2020, BoucheMEGAFLOW2025}.
With large numbers of QSO-galaxy pairs, it is possible to look for correlations between galaxy and absorber properties. The extent and strength of \mgii\ absorption depends on stellar mass, star formation rate, specific star formation rate (sSFR$=$SFR/\mgal) or colour of the galaxies \citep[e.g.][]{ChenHW2010a,ChenWild2010b,Bordoloi2011,Nielson2013,Lundgren2021,Huang2021, Dutta2020}. The results are consistent with more massive galaxies living in more massive halos with correspondingly greater extents of \mgii clouds, while the secondary trend with ongoing star formation rate could be due to either greater outflows or greater gas feeding rates around higher sSFR galaxies.  

% MgII observations - stacked 

Another approach exploiting the advent of large scale galaxy surveys involves looking for an excess \mgii absorption around known galaxies, either via cross correlation with known absorbers or via stacking of QSO sightlines. These approaches can detect excess \mgii absorption out to 10-20\,Mpc from massive ($\log_{10}\mgal/\modot>10$) galaxies, with the mean \mgii equivalent width declining with radius as ${\rm W_{MgII}} \propto R^{a}$, with $a\sim2$, and the exact value dependent on sample \citep{Zhu2014,PerezRafols2015,Kauffmann2017,Lan2018,Wu2025,ChenZ_DESIMgII_2025}. These methods are sensitive to weak absorbers ($\lesssim0.3$\AA) to which individual absorber studies are less sensitive, but also tend to probe gas around higher mass galaxies than absorber-targeted samples, due to the galaxy flux-limited selection. While no mass dependence in the radial profile of \mgii is found at high masses, galaxies with $\log_{10}\mgal/\modot<10$ show significantly lower ${\rm W_{MgII}}$ at a given impact parameter, and a smaller extent is detected \citep{ChenZ_DESIMgII_2025}. Both SFR and orientation between the galaxy major axis and the absorber are also found to impact the strength of \mgii absorption observed, as seen in galaxy-absorber pair studies \citep{Lan2014,Lan2018,Anand2021,Wu2025,ChenZ_DESIMgII_2025}.

In summary, it is well known that galaxy properties such as stellar mass and star formation rate are correlated with cool CGM properties for a wide range of datasets, but revealing the physical mechanisms responsible for the correlations is challenging. Here we add a new dimension, investigating for the first time the connection between galaxies which underwent a recent burst of star formation that has rapidly quenched (post-starburst galaxies) and the \mgii absorption in the surrounding CGM. This builds on previous studies of Ly$\alpha$ and high ionisation metal lines along small numbers of sightlines close to low redshift starburst and post-starburst galaxies \citep[COS-burst:][]{Borthakur2013,Heckman2017}, and shows us how galaxies and CGM vary together on timescales of several hundred Myr. Section \ref{sec:data} details the datasets used, Section \ref{sec:methods} outlines the methodology, Section \ref{sec:results} presents the results, and Section \ref{sec:discussion} discusses the interpretation. Conclusions are summarised in Section \ref{sec:summary}. Throughout this paper, we assume a flat, $\Lambda$ cold dark matter model with $H_0 = 68\,{\rm km\,s}^{-1}\,\rm{Mpc}^{-1}$, $\Omega_m = 0.3$ and $\Omega_\Lambda = 0.7$.

%%%%%%%%%%%%%%%%%%%%%%%%%%%%%%%%%%%%%
\section{Data}\label{sec:data}
%%%%%%%%%%%%%%%%%%%%%%%%%%%%%%%%%%%%%

All data used in this study are from the Sloan Digital Sky Survey (SDSS), which uses the 2.5\,m aperture Sloan Foundation telescope located at the Apache Point Observatory in New Mexico \citet{gunn25TelescopeSloan2006}. Over several decades, the SDSS has imaged 1/3 of the sky in 5 optical and NIR bands ($u$, $g$, $r$, $i$ and $z$) and obtained optical and NIR spectroscopic follow-up for nearly 5 million objects, collecting data for a wide range of science goals. In this paper we make use of the spectroscopic QSO catalogue \citep{SchneiderQSO2010} from SDSS data release 7 \citep[DR7,][]{SDSSDR7_2009} and the Baryon Oscillation Spectroscopic Survey (BOSS) CMASS galaxy spectroscopic catalogue from the SDSS data release 12 \citep[DR12,][]{SDSSDR12_2015}.

%-------------------------------------
\subsection{The SDSS DR7 quasar sample}\label{sec:data_qsos}
%-------------------------------------
The QSO catalogue used in this paper includes 105,783 spectroscopically confirmed QSOs brighter than $i$-band absolute magnitude $M_i=-22.0$ from the SDSS DR7 \citep{SchneiderQSO2010}. While the catalogue contains QSOs with $0.065<z<5.46$, we only use QSOs with $0.5 \lesssim z \lesssim 3$ (see Section \ref{sec:data_pairs}). The spectra cover a wavelength range of 3800-9200\AA\ with a spectral resolution of $\lambda/\Delta\lambda=1850-2200$ \citep{SDSSDR7_2009}. Although a larger quasar sample was available from SDSS Data Release 16 (DR16), for the purposes of this paper the higher signal-to-noise ratios (SNR) of the DR7 dataset was preferred.

%-------------------------------------
\subsection{The CMASS galaxy sample}\label{sec:data_gals}
%-------------------------------------

The SDSS BOSS survey \citep{DawsonBOSS2013} obtained spectra of galaxies and QSOs with the re-built SDSS spectrographs \citep{Smee_BOSSspectrographs2013}, which have a wavelength range of 3600-10,000\AA. The CMASS sub-sample of the BOSS survey was designed to target galaxies over a nominal redshift range of $0.4\lesssim z\lesssim0.7$ with a colour selection to make the sample approximately volume-limited in stellar mass \citep{Reid2016}. The CMASS DR12 large scale structure galaxy catalogue covers about 10,400\,deg$^2$ and contains spectra for 849,637 unique galaxies\footnote{\url{https://data.sdss.org/sas/dr12/boss/lss/}; CMASS ``dr12v5'' north and south}. We restricted the sample to $0.35<z<0.8$, corresponding to the minimum redshift for observing the \mgii doublet and the maximum redshift our galaxy spectral classification method can be used in BOSS spectra, which gives 830,402 galaxies.

The stellar masses for the galaxies were taken from the Granada group value added catalogue \citep{MonteraDorta2016}, using the version that fits for dust attenuation and uses a `wide-star-formation' that allows an extended star formation history\footnote{\url{https://www.sdss4.org/dr12/spectro/galaxy_granada/}, filename granada\_fsps\_krou\_wideform\_dust-DR12-boss.fits.gz}.   In this catalogue, stellar masses were calculated from fitting FSPS stellar population synthesis models \citep{Conroy_FSPS2009} to the $ugriz$ extinction corrected model magnitudes, using the BOSS spectroscopic redshifts. In this paper we chose to use the median posterior stellar mass value. Matching the CMASS DR12 ``v5'' north and south catalogues to this stellar mass catalogue resulted in 814,494 galaxies.

\begin{table*}
    \centering
    \caption{The number of galaxy-quasar pairs in bins of impact parameter (transverse distance between the galaxy and quasar sightline), for all, post-starburst, star-forming and quiescent galaxies. The mean impact parameter is calculated over all pairs in the bin. Note that we classify, but do not use, green-valley galaxies, and therefore the numbers of pairs in individual galaxy types does not add up to the total number of pairs.}
     \begin{tabular}{c|c|c|c|c|c}
       Impact Parameter & Mean Impact & \multicolumn{4}{c}{Number of pairs} \\
                      (kpc)         & Parameter (kpc)  & all & post-starburst & star-forming & quiescent \\
         \hline
         0 - 71 & 51 & 115 & 9 & 24 & 75\\
         71 - 141 & 112 & 420 & 13 & 89 & 258\\
         141 - 282 & 219 & 1854 & 39 & 329 & 1188\\
         282 - 566 & 442 & 7577 & 202 & 1318 & 4825\\
         566 - 1131 & 883 & 30260 & 863 & 5411 & 19046\\
         1131 - 2262 & 1757 & 122451 & 3320 & 22176 & 77323\\
         2262 - 4525 & 3421 & 488348 & 13418 & 87829 & 309800\\
         4525 - 9051 & 7800 & 1947357 & 53750 & 350047 & 1233515\\
         \hline
    \end{tabular}
    \label{tab:pairs}
\end{table*}

%-------------------------------------
\subsubsection{Galaxy Classifications}\label{sec:data_specclass}
%-------------------------------------

In order to separate galaxies based on their recent star formation history we fit the galaxy spectra using the mean-field independent component analysis (MFICA) galaxy templates of \citet{Krishna2025}. These 7 templates are based on a training sample of SDSS DR7 galaxy spectra which  encompasses most of the variation observed in the galaxy spectra in our sample. The decomposition uses the rest-frame wavelength range 3300-5100\AA\ and SDSS DR7 galaxies with redshifts $0.16<z<0.38$. The first three components predominantly reconstruct the stellar continuum of the galaxy spectra, each representing the contribution of stellar populations of different ages: young (``OB''), intermediate (``AF'') and old (``K''). A further three components reproduce the nebular emission line strengths, and a final one adjusts the width of the emission line. 

The relative light fractions of the three stellar continuum components provide a data-driven representation of the recent star formation history of the galaxy. The stellar continuum  of quiescent galaxies is dominated by light from long-lived, low mass stars, and therefore these galaxies have high K component fractions\footnote{The slightly lower redshift of the template galaxies compared to the CMASS sample means the old component will be very slightly too old. However, this difference is expected to be negligible given the very slow change in spectral shape with age beyond 4\,Gyr. As we use the components in a relative sense for classification only this does not impact our results.}. Star-forming galaxies are made up of varying combinations of the three components, dependent on their sSFR, with starburst galaxies having the largest OB component fraction due to their short-lived burst of star-formation. Post-starburst galaxies, which have recently experienced a starburst that has rapidly quenched, can be identified by the strong contribution of intermediate lifetime A and F stars to their light from the recent starburst. These have a higher AF component fraction than normal star-forming galaxies at fixed K component fraction. The properties of high mass CMASS-selected post-starburst galaxies have been presented recently \citep{Suess2022, Bezanson2022, Setton2025}, however, their CGM properties have not yet been investigated.

%-------------------------------------
\subsection{Galaxy-quasar pairs}\label{sec:data_pairs}
%-------------------------------------

\begin{figure*}
	\includegraphics[width=\columnwidth, trim={0.2cm 0 0 0}]{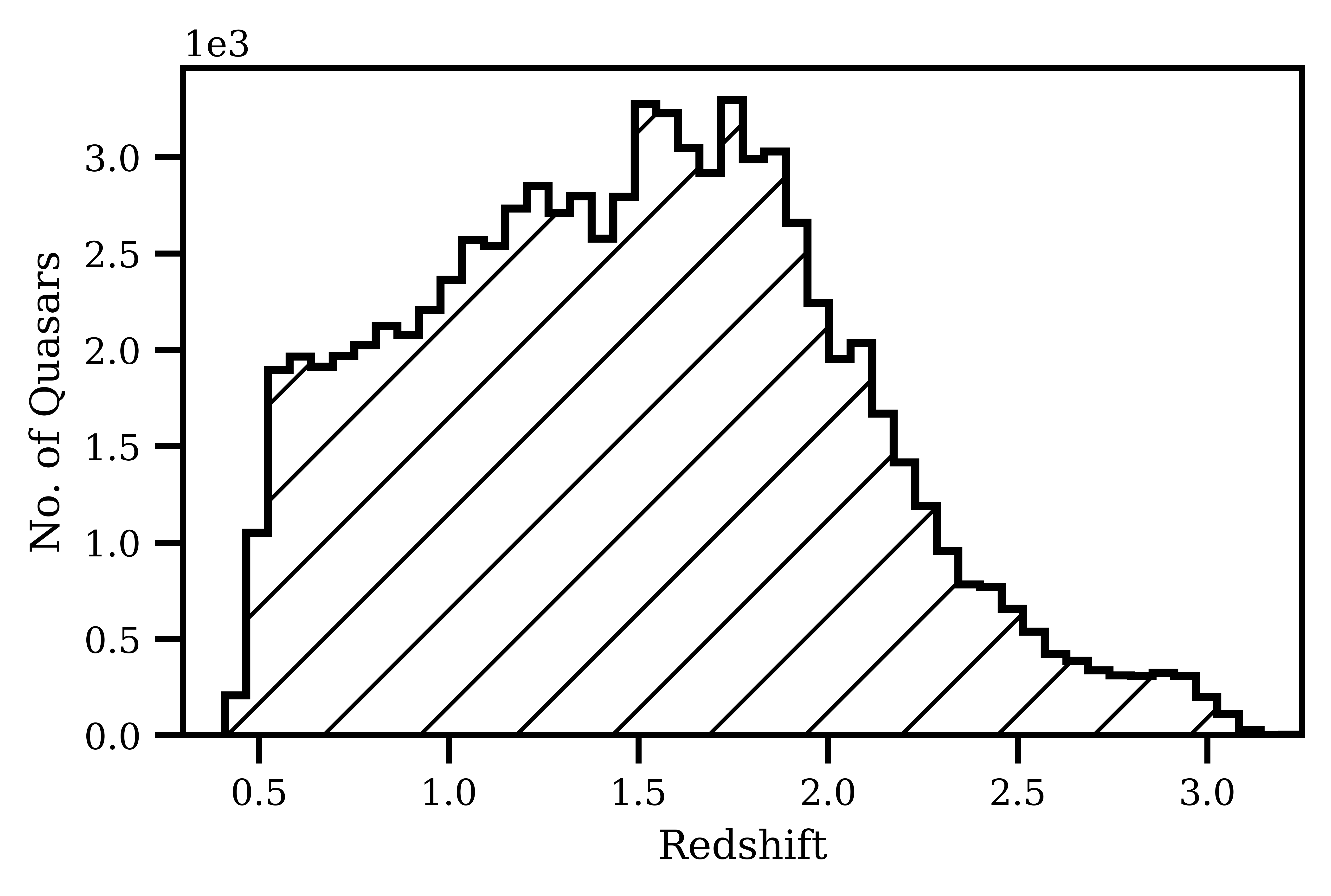}
    \includegraphics[width=\columnwidth, trim={0 0 0 0}]{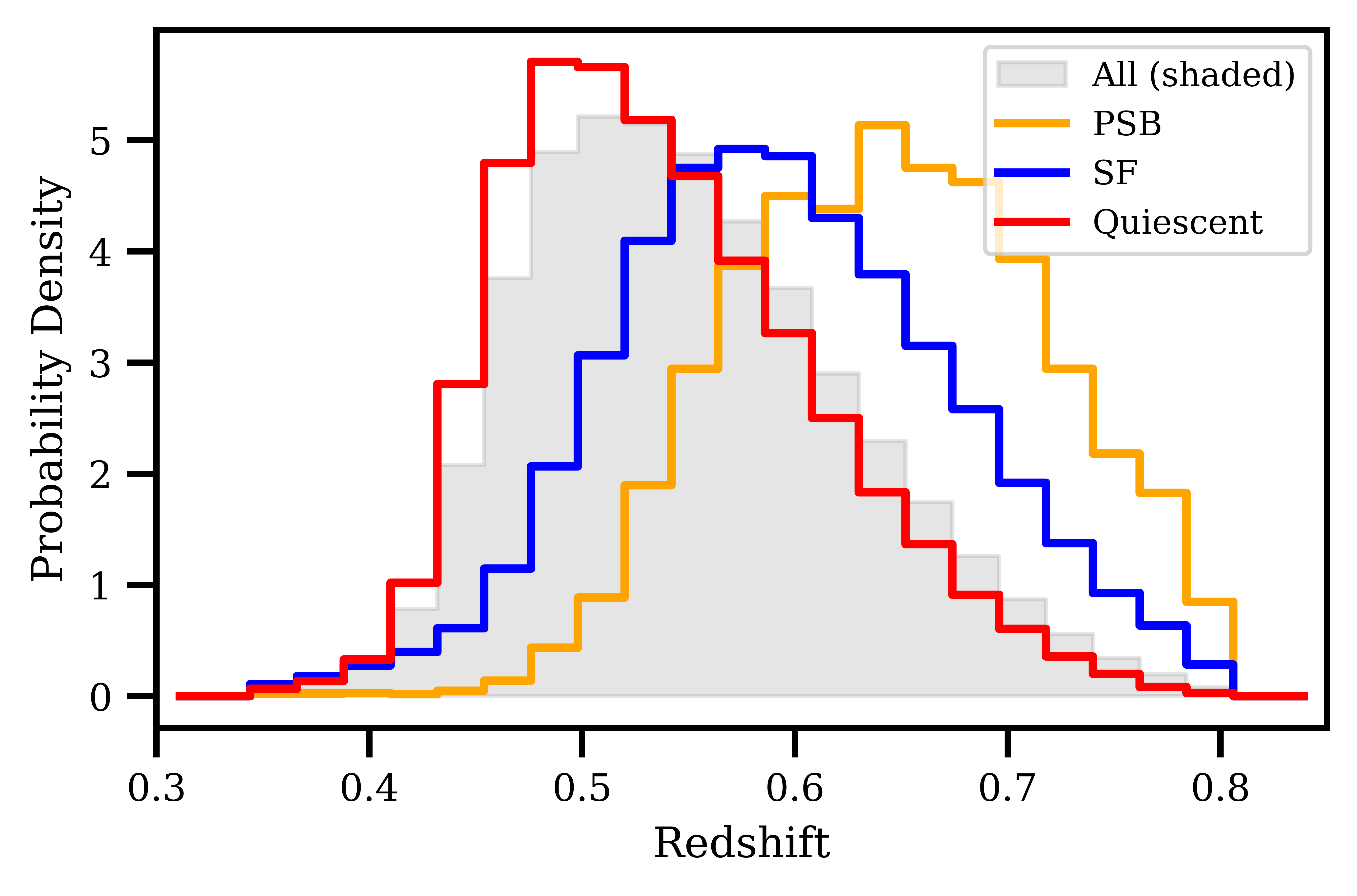}
    \caption{The redshift distribution of the unique SDSS DR7 quasars (left panel) and unique SDSS DR12 CMASS galaxies (right panel) that make up the galaxy-quasar pairs studied in this paper. The galaxy histograms show the whole sample (grey filled), quiescent (red line), star-forming (blue line) and post-starburst (orange line) sub-samples. Each galaxy and each QSO may contribute to multiple pairs in our spectral stacks. }
    \label{fig:redshifts}
\end{figure*}

\begin{figure*}
	\includegraphics[width=\columnwidth, trim={0.1 0 0.2cm 0}]{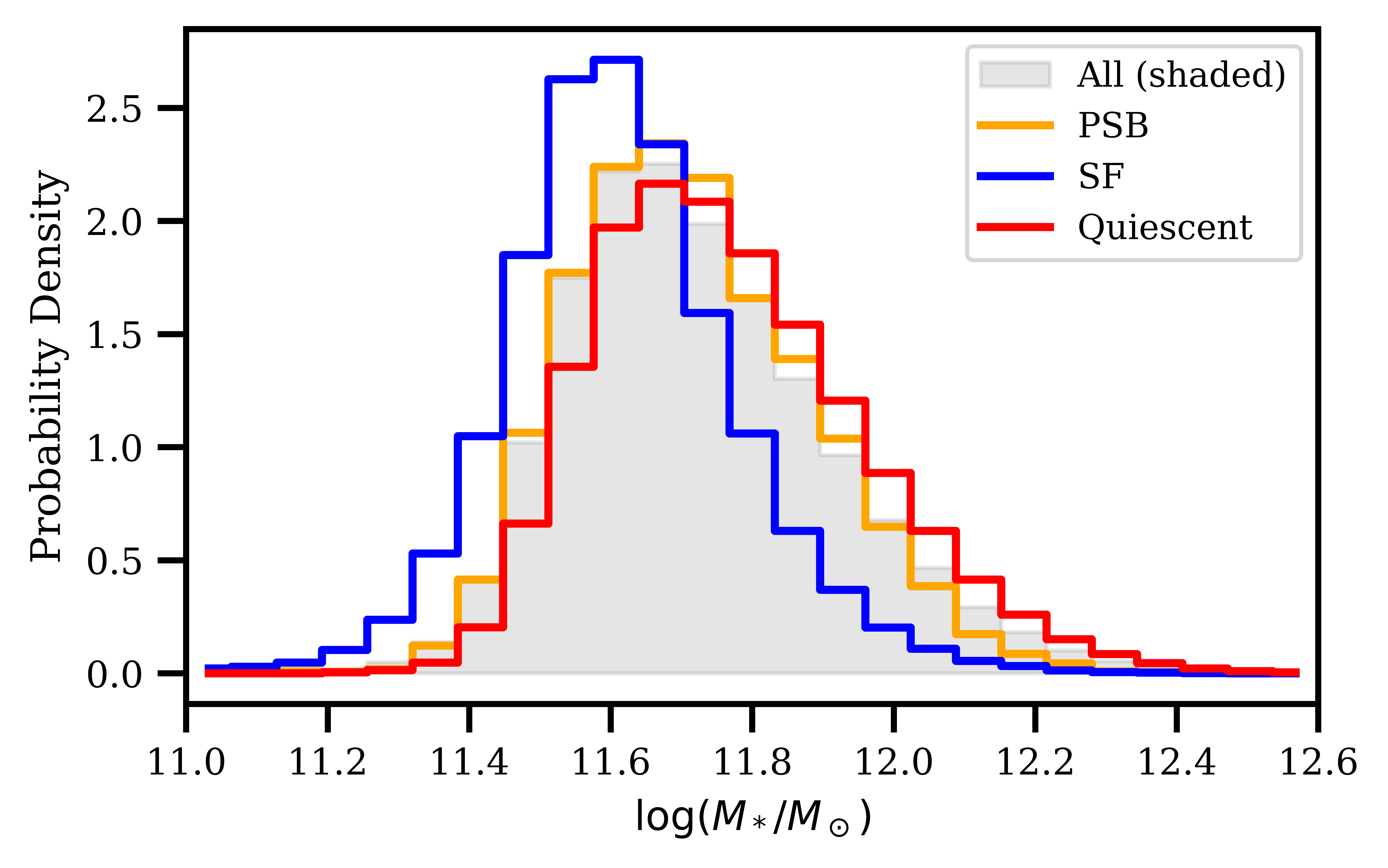}
    \includegraphics[width=\columnwidth, trim={0 0 0.2cm 0}]{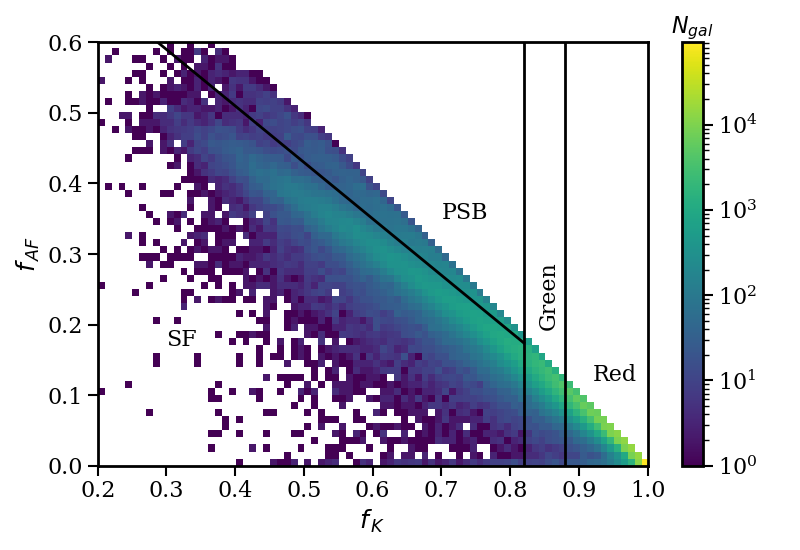}
    \caption{\emph{Left:} The stellar mass distribution of the unique SDSS DR12 CMASS galaxies that are part of the galaxy-quasar pairs studied in this paper. The histograms show the whole sample (grey filled), quiescent (red line), star-forming (blue line) and post-starburst (orange line) sub-samples. As noted in the text, there is significant disagreement in stellar masses between different CMASS value added catalogues, and therefore these histograms should be taken as indicative only.  \emph{Right:} The MFICA derived K and AF component fractions of the galaxy spectra, with colour scale indicating the logarithmic number of galaxies. The K  component fraction represents the fraction of galaxy light between 3300-5100\AA\ contributed by long-lived, low-mass stars. The AF component fraction represents the fraction of light contributed by intermediate mass A and F stars. Based on these two light fractions, the galaxy is classified as star forming (``SF''), quiescent (``Red'') or post starburst (``PSB'') as indicated by the demarcation lines. Green valley galaxies, lying between the quiescent and star-forming populations, are excluded from our analysis as they were found to contain significant contamination from the quiescent galaxy population in the relatively low SNR CMASS spectra. }
    \label{fig:galaxy_properties}
\end{figure*}

The galaxy and quasar samples described in Sections \ref{sec:data_gals} and \ref{sec:data_qsos} were cross-matched to form galaxy-quasar line-of-sight pairs by calculating the angular separation on the sky between the galaxy and the quasar using the right ascension and declination of both objects. The angular separation was converted into a transverse distance (kpc) using the angular diameter distance to the redshift of the galaxy, with a maximum transverse distance limit of 9.051\,Mpc. This separation between galaxy and quasar sightline is termed the ``impact parameter''.

We removed galaxy-quasar pairs where any of the following were true:
\begin{itemize}
    \item the quasar redshift was lower than the galaxy redshift; 
    \item the galaxy-quasar pair had a velocity difference such that the quasar redshift was within $0.04\,c$ of the galaxy, to ensure that the absorption line systems associated with quasar outflows were excluded;
    \item the galaxy-quasar pair had a redshift difference such that the galaxy's \mgii absorption feature fell in the Lyman-$\alpha$ forest region of the quasar spectrum (i.e. $<1215.67$\AA\ in the rest-frame of the quasar);
    \item the galaxy-quasar pair had a redshift difference such that the galaxy's \mgii absorption feature fell in strong emission line regions of the quasar spectrum (specifically 30\AA\ centred on \civ$\lambda1549$  and 20\AA\ centred on \ciii$\lambda1909$). 
\end{itemize} 
A small number of pairs (7) with low SNR or strong visually identified absorbers which contributed disproportionately to the final weighted stacked spectra were additionally excluded.  This resulted in a total of 2,598,382 quasar-galaxy pairs, including 470,116 unique galaxies and 82,781 unique quasars. These pairs were divided into bins based on their transverse distance (i.e. impact parameter) from one another and the total number of pairs in each bin is shown in the second column of Table \ref{tab:pairs}. 

The redshift distribution of the unique galaxies and quasars that make up the galaxy-quasar pairs are shown as a grey shaded histogram in Fig.~\ref{fig:redshifts}, showing that our analysis probes the CGM of galaxies at $0.4\lesssim z\lesssim 0.8$. Note that most galaxies and quasars contribute to multiple pairs. The left panel of Fig.~\ref{fig:galaxy_properties} presents the stellar mass distribution of the galaxies in the final pair sample as a grey shaded histogram. We note that the stellar masses in the Granada value added catalogue are, on average, 0.25-0.5\,dex larger than in other value added catalogues for the CMASS dataset, showing that reported stellar masses based on photometry at these redshifts are significantly dependent on prior assumptions and methodology. For this reason, as well as number density considerations, we conclude only that the galaxies are more massive than the characteristic stellar mass of galaxies at these redshifts of $\log_{10}({\rm M/M_\odot})=10.96\pm0.1$ \citep{Weaver2023}.

The right panel of Fig.~\ref{fig:galaxy_properties}, shows the AF vs. K component fractions of the galaxy spectra. The galaxies are split into star-forming, quiescent, green-valley and post-starburst, using demarcation lines presented in \citet{Krishna2025}. Note that the component fractions are derived from normalised spectra, and therefore they represent mass-normalised quantities (i.e. our ``star-forming'' sample is comparable to samples with a high specific star formation rate, rather than absolute star formation rate). The green-valley sample was excluded from further analysis in this paper, as there was no clear difference in the stacked spectra of this sub-class compared to quiescent galaxies, indicating that this sample suffered from high contamination from the dominant quiescent galaxy sample scattering into it, due to the relatively low SNR of the galaxy spectra used in this paper. Due to the very high stellar mass of the galaxy sample, the vast majority of the galaxies are classified quiescent (i.e. Luminous Red Galaxies, LRGs, 75\%), however, a significant fraction of the sample are star-forming (21\%) or post-starburst (3\%). This low fraction of post-starburst galaxies matches expectations from the literature at these redshifts and masses, using similar selection methods \citep[e.g.][]{Rowlands2018}. 

The number of pairs in each spectral class and impact parameter bin are included in the final 3 columns of Table \ref{tab:pairs}. Stacked spectra for galaxies in each of the three samples are shown Appendix \ref{app:stacks}. The redshift and stellar mass distributions of the three sub-samples are shown as coloured histograms in Figs.~\ref{fig:redshifts} and \ref{fig:galaxy_properties}. There is a difference in the redshift distribution of galaxies with different stellar populations, which is caused by the colour-based targetting of the CMASS sample, which was aiming for a mass-limited sample \citep{Reid2016}. The difference in the stellar masses of the samples is very small, as expected. We investigate the possible impact of the different redshift and stellar mass distributions on our results in Appendix \ref{app:systematics}.

%%%%%%%%%%%%%%%%%%%%%%%%%%%%%%%%%%%%%
\section{Method}\label{sec:methods}
%%%%%%%%%%%%%%%%%%%%%%%%%%%%%%%%%%%%%
In this section we detail the procedures used to remove the quasar continuum from the background quasar spectra, and stack the spectra at the redshift of the galaxies to create high SNR stacks of the \mgii absorption line in the CGM around the galaxies. We then present the measurement of the galaxy rest-frame equivalent width of the \mgii absorption lines (${\rm W_{MgII}}$). 

\subsection{Removing the quasar continuum}

To remove the continuum associated with the quasar emission in the quasar spectra, we used a series of median filters and sigma-clipping. Before applying the filtering, we first masked a 20\AA\ region centred on the \mgii doublet in the galaxy rest-frame. This ensures that the median filter does not remove the signal we are trying to detect. Strong skylines in the observed-frame were also masked. We then applied a running median filter of 45 pixels, including across the masked regions (i.e. reducing the number of contributing pixels) and the quasar flux and error array were divided by this running median continuum. Next, we masked all pixels with absolute flux/error $>2.5$, replacing masked pixels with the median filter value. We repeated the process of median filtering and masking outliers two more times to ensure that the number of outlier pixels was very low. We ran two final iterations with a filter width of 15 and 65 pixels, which reduced any residual continuum close to the \mgii doublet. This procedure was arrived at through trial and error by creating stacks of known \mgii\ absorption systems to ensure that the continuum remained flat around the absorber doublet. 

\subsection{Stacking} 

To create the stacked spectra, the individual spectra were shifted to the absorber rest-frame without rebinning and the region between 2750\AA\ and 2850\AA\ was extracted. We created a weighted stack of continuum divided spectra for all galaxy-quasar pairs in the impact parameter bins given in Table \ref{tab:pairs}.  We used a single weight for each spectrum, calculated from the mean of the inverse squared error array within the 100\AA\ centred on the \mgii doublet. This ensured that spectra with higher noise contributed proportionally less to the final stack, thereby reducing the overall noise in the stacked spectrum. At the end of the stacking process, the stacked flux was normalised by dividing through by the total of the weights contributing at each wavelength, producing the final stacked spectrum. 

\subsection{Equivalent widths}

To evaluate the strength of the \mgii doublet the equivalent width was computed from the stacked spectra by directly integrating the flux over the region of the spectrum between 2791-2808\AA. Both \mgii doublet lines were included in the integration, and no Voigt-profile modelling was performed. Errors were estimated from bootstrap resampling the galaxy-quasar pairs with replacement, repeating the stacking procedure and the equivalent width measurement 1000 times. The error on the equivalent width values was taken to be the standard deviation from these 1000 measurements. 

%%%%%%%%%%%%%%%%%%%%%%%%%%%%%%%%%%%%%
\section{Results}\label{sec:results}
%%%%%%%%%%%%%%%%%%%%%%%%%%%%%%%%%%%%%

\begin{figure*}
    \centering
    \includegraphics[width=\textwidth]{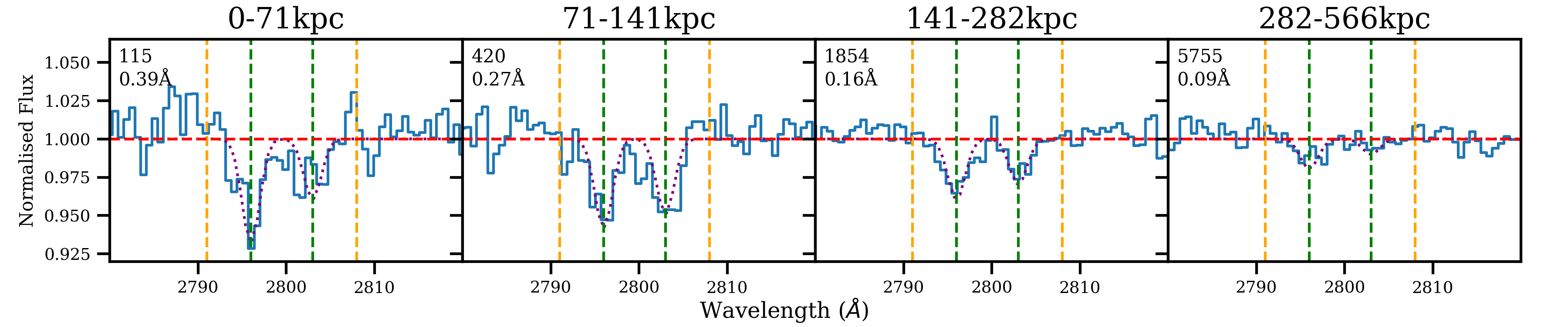}
    \includegraphics[width=\textwidth]{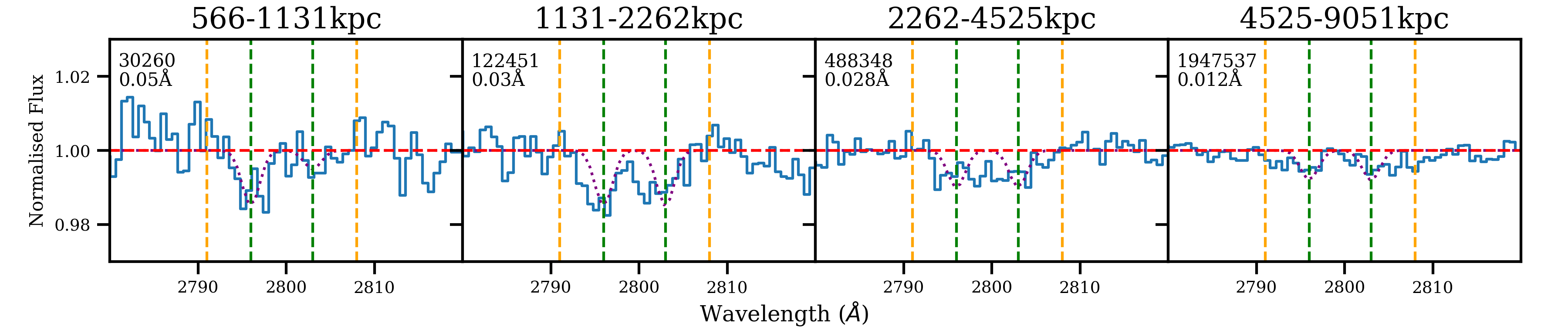}
    \caption{Stacked spectra around the \mgii absorption doublet in the rest-frame of the galaxies for the inner four (top) and outer four (bottom) impact parameter bins as given in the title of each subplot. Note the change in y-axis scaling between the upper and lower panels. The green dashed line shows the expected positions of the \mgii absorption doublet, and the orange dashed lines show the wavelengths between which the equivalent width is measured. For visualisation purposes only, we also show a double Gaussian profile fit to the \mgii absorption lines (purple dotted lines). The top left of each panel gives the number of pairs included in the stack, and the measured \mgii equivalent width of each stack. An absorption signature is seen at all impact parameters. }
    \label{fig:allstacks}
\end{figure*}

\begin{figure*}
   \centering
    \includegraphics[width=\textwidth]{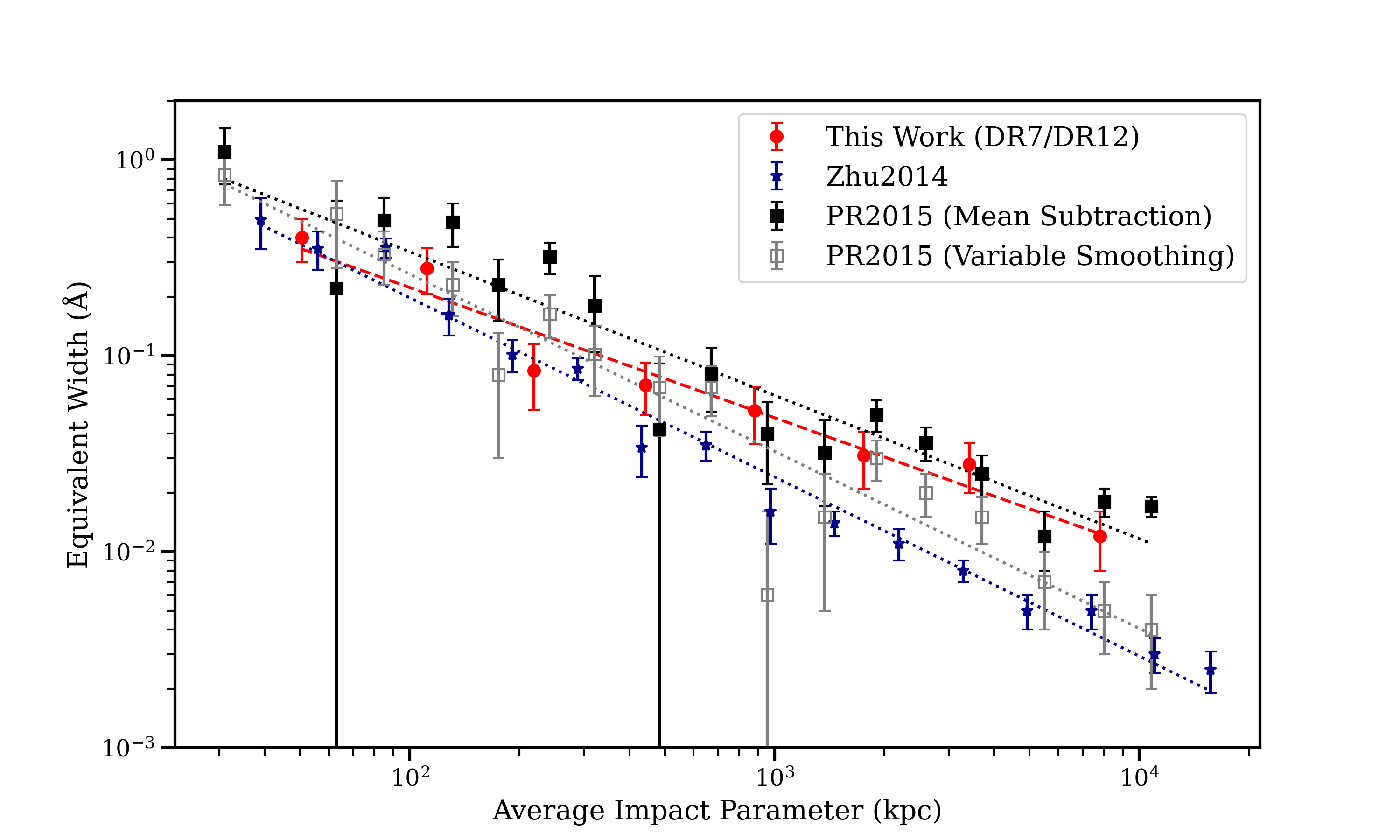}
    \caption{The equivalent width of the \mgii absorption doublet as a function of average impact parameter for our work (red points), compared to results from the literature. Error bars are from a bootstrap analysis as described in the text. The green points are from \citet{Zhu2014} and the black and grey points are from \citet{PerezRafols2015}, the black points show their variable smoothing method for removing the quasar continuum and the grey points show their mean subtraction method. Dotted lines show least squares linear fits to each dataset. In agreement with past results, we find that the strength of \mgii absorption in the CGM of galaxies decreases with impact parameter.}
    \label{fig:eqw_b}
\end{figure*}

\begin{table*}
    \centering
    \caption{The equivalent width of the \mgii doublet as a function of distance (impact parameter) from the galaxy for all, post-starburst, star-forming and quiescent galaxy samples. Errors are from a bootstrap analysis. }
        \begin{tabular}{c|c|c|c|c}
          Impact Parameter (kpc) &   \multicolumn{4}{c}{Equivalent width (\AA)} \\
          & All & post-starburst & star-forming & quiescent \\
         \hline
   0 - 71           &  0.39 ± 0.10 & 1.37 ± 0.24 & 0.49 ± 0.04 & 0.18 ± 0.01\\
   71 - 141       &    0.27 ± 0.07 & 0.59 ± 0.06 & 0.28 ± 0.02 & 0.15 ± 0.01\\
   141 - 282     &     0.16 ± 0.05 & 0.37 ± 0.04 & 0.13 ± 0.01 & 0.078 ± 0.009\\
   282 - 566     &     0.09 ± 0.02 & 0.08 ± 0.01 & 0.0679 ± 0.008 & 0.048 ± 0.008\\
   566 - 1131   &      0.05 ± 0.01 & 0.035 ± 0.005 & 0.037 ± 0.004 & 0.043 ± 0.007\\
   1131 - 2262 &      0.03 ± 0.01 & 0.027 ± 0.004 & 0.029 ± 0.003 & 0.039 ± 0.090\\
   2262 - 4525 &      0.028 ± 0.008 & 0.025 ± 0.003 & 0.033 ± 0.002 & 0.035 ± 0.008\\
   4525 - 9051 &       0.012 ± 0.004 & 0.017 ± 0.002 & 0.030 ± 0.001 & 0.025 ± 0.007\\
         \hline
    \end{tabular}
    \label{tab:eqw}
\end{table*}

\begin{figure*}
    \centering
    \includegraphics[width=\textwidth]{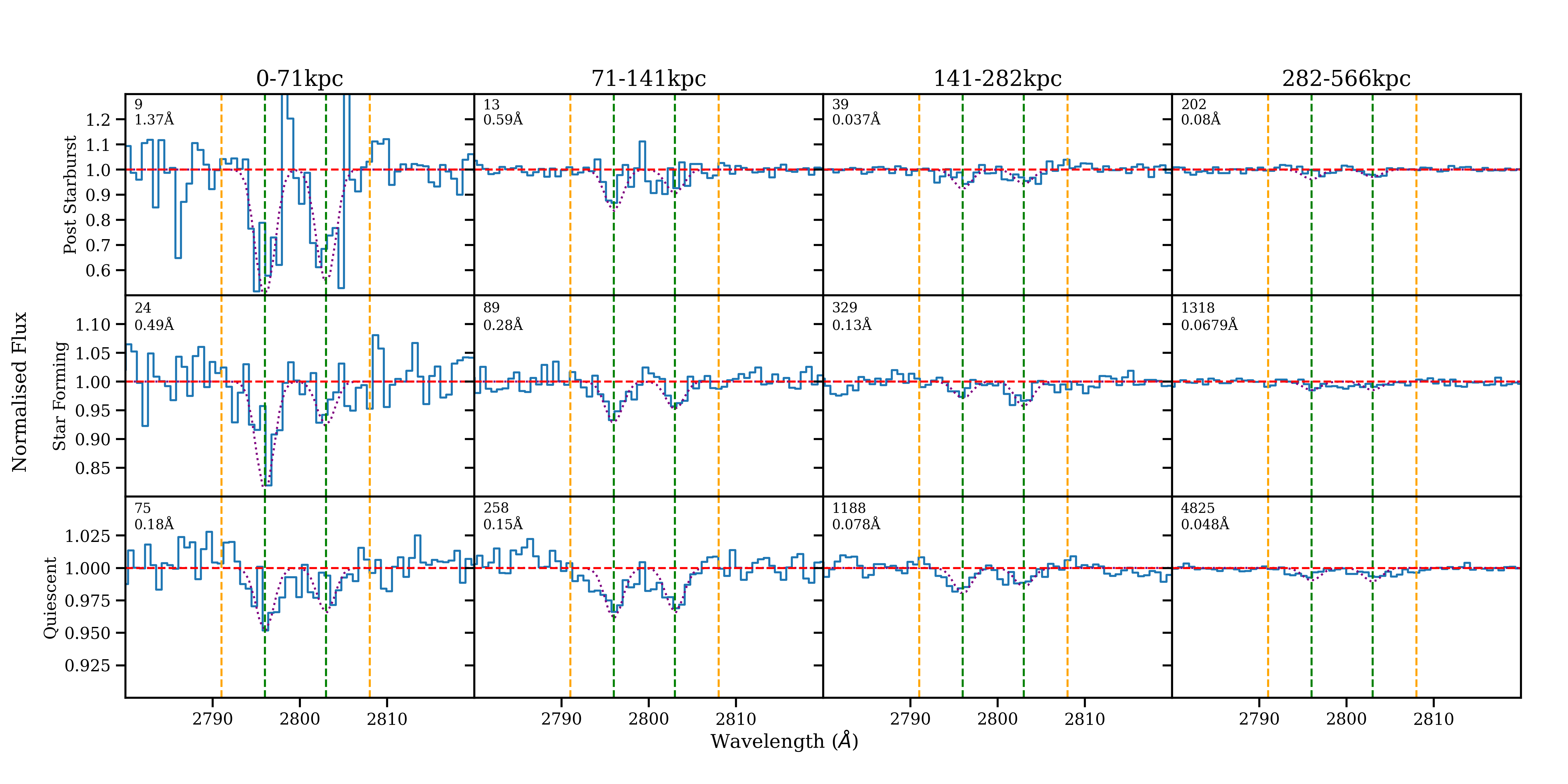}
    \caption{Stacked spectra around the \mgii absorption doublet in the rest-frame of the galaxies for the inner four impact parameter bins, for galaxies of different spectral types: post-starburst (top), star-forming (centre) and quiescent (bottom). Note the change in y-axis scaling between rows. The dashed and dotted lines are the same as in Fig.~\ref{fig:allstacks}. The top left of each panel gives the number of pairs included in the stack, and the measured \mgii equivalent width of each stack}. The absorption signal decreases with impact parameter for all types of galaxy, and is notably stronger in the post-starburst galaxies. The quiescent galaxies have the weakest signal.
    \label{fig:classstacks}
  \end{figure*}

  \begin{figure*}
    \centering
    \includegraphics[width=\textwidth]{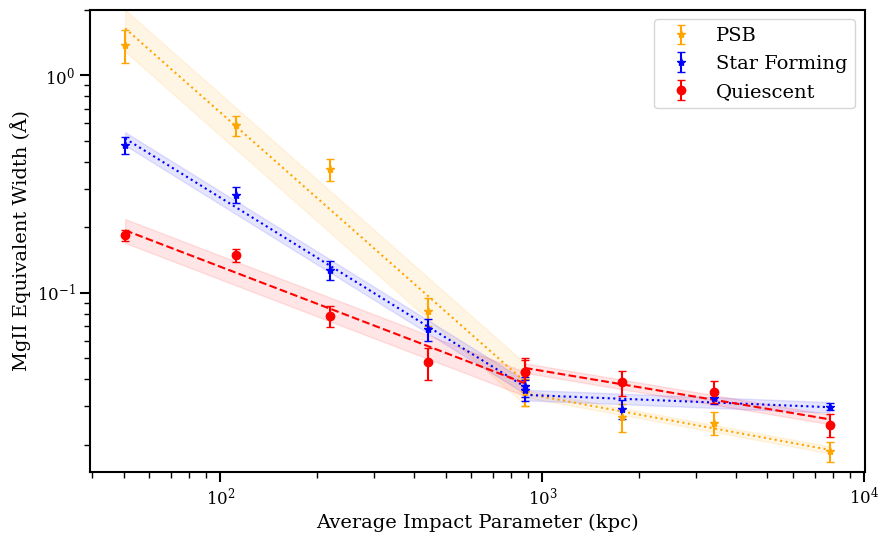}
    \caption{The equivalent width of the \mgii absorption doublet as a function of average impact parameter for galaxies of different spectral types: post-starburst (orange), star-forming (blue) and quiescent (red). Lines indicate least squares linear fits, independently below and above 1\,Mpc, with error bounds calculated from the root-mean-square scatter of data points around the fitted line. The post-starburst galaxies have much stronger CGM absorption within 1\,Mpc than the quiescent or star-forming galaxies. The steep profiles of the post-starburst and star-forming galaxies flatten beyond 1\,Mpc. }
    \label{fig:eqw_b_class}
    \end{figure*}

    \begin{table*}
    \centering
    \caption{The linear best fit parameters to the relation between $\log_{10}$ of \mgii\ absorption doublet equivalent width (in \AA) and $\log_{10}$ of impact parameter (in kpc) for all, post-starburst, star-forming and quiescent galaxy samples within and outwith 1\,Mpc. }
        \begin{tabular}{c|c|c|c|c|c|c|c|c}
          & \multicolumn{2}{c|}{All} & \multicolumn{2}{c|}{post-starburst} & \multicolumn{2}{c|}{star-forming} & \multicolumn{2}{c}{quiescent} \\
          & gradient & intercept & gradient & intercept & gradient & intercept & gradient & intercept \\
         \hline
   Within 1Mpc           &  -0.7 & 0.8     & -1.3 & 2.5       & -0.9   &  1.3       & -0.6 &  0.2\\
   Outwith 1Mpc        &  -0.7 & 0.8     & -0.3 & -0.5      & -0.07 & -1.3      &  -0.3 & -0.6\\
         \hline
    \end{tabular}
    \label{tab:linearbestfit}
\end{table*}
  
    Fig.~\ref{fig:allstacks} shows the stacked spectra around the \mgii absorption line associated with the total galaxy sample, for all eight bins of impact parameter. The equivalent width of the \mgii doublet is calculated by summing between the yellow dashed lines. For visualisation purposes only, we also show a double Gaussian profile least-squares fit to the \mgii absorption lines, with free amplitude but fixed wavelengths and widths. Fig.~\ref{fig:eqw_b} shows the equivalent width vs. average (mean) impact parameter of our results compared to results from \citet{PerezRafols2015} and \citet{Zhu2014}, together with least squares linear fits. A clear decline in equivalent width with distance into the CGM is observable in both plots, with the amplitude and decline rate broadly in agreement with previous results from the literature. The equivalent width measurements and errors are provided in the left hand column of Table \ref{tab:eqw} and the parameters of the linear fit to our dataset is given in Table \ref{tab:linearbestfit}.  

    To test the robustness of our results at large impact parameters, where small equivalent widths are measured, we randomised the redshifts of the galaxies and repeated the spectral stacks, thereby stacking the QSO spectra at randomised wavelength positions. Each galaxy in each pair recieved a randomised redshift selected from another pair in the dataset, all impact parameters were recalculated, and the spectral stacks repeated with the new samples. Equivalent widths of between -0.0054 and 0.0052\AA\ were measured, with equal numbers of negative and positive values and bootstrap errors indicated the measurements were consistent with zero. This demonstrates that we are significantly detecting absorption even at the largest distances probed by our results in Fig.~\ref{fig:allstacks} and that the bootstrap errors for bins with large numbers are robust. 

    Fig.~\ref{fig:classstacks} shows the stacked spectra around the \mgii absorption line for the inner four bins of impact parameter and each of the galaxy spectral types: star-forming, quiescent and post-starburst. Equivalent width measurements and errors are provided in Table \ref{tab:eqw}. Fig.~\ref{fig:eqw_b_class} shows the relationship between the \mgii equivalent width and impact parameter for the three galaxy subsamples, over the full range in impact parameter. For each galaxy type, we perform independent least squares linear fits with free slope and amplitude to the $\log_{10}({\rm W_{MgII}})$ vs. $\log_{10}(b)$ data, where ${\rm W_{MgII}}$ is the equivalent width of the \mgii\ doublet in \AA, and $b$ is the impact parameter in kpc. Two fits are performed separately to the inner 5 and outer 4 data points, thus splitting the data at $\sim$1\,Mpc and shown as lines on  Fig.~\ref{fig:eqw_b_class}. Error bounds are calculated from the sum of the squared residuals of the least squares fit. Table \ref{tab:linearbestfit} provides the linear fit parameters. We see that star-forming galaxies exhibit significantly stronger \mgii absorption in their CGM compared to quiescent galaxies. However, post-starburst galaxies exhibit by far the strongest absorption in the inner 3 bins ($<$282\,kpc). Beyond 1\,Mpc, all subsamples of galaxies exhibit similar levels of absorption, and the relationship between equivalent width and impact parameter flattens off. 

    In Appendix \ref{app:systematics} we show Fig.~\ref{fig:eqw_b_class} with the quiescent and star-forming samples split by redshift, to demonstrate that the difference is not driven by the slightly different redshift distributions of the three samples (Fig.~\ref{fig:redshifts}). While it is common in the literature to stack spectra in radial impact parameter bins normalised by an estimated halo virial radius, we found that this was not possible for our sample, due to the large errors on stellar mass discussed above, as well as the stellar mass - halo mass relation being relatively unconstrained at these high stellar masses. To verify that our results are not expected to change significantly should we be able to normalise by halo virial radius, which is determined via a scaling of stellar mass, we show in Fig.~\ref{fig:eqw_b_class_mass} that the absorption strength is independent of stellar mass in all sub-samples.

\section{Discussion}\label{sec:discussion}

The decreasing strength of \mgii absorption with distance from the galaxy, as seen in Fig.~\ref{fig:eqw_b}, is consistent with the well-documented results in the literature from both individual absorber \citep[e.g.][]{ChenHW2010a,Nielson2013,Lundgren2021} and stacked sightline studies \citep[e.g.][]{Bordoloi2011,Zhu2014, PerezRafols2015,ChenZ_DESIMgII_2025}. Clearly the most significant factor influencing \mgii absorption strength is the distance from the galaxy. Although variations of absorber strength with stellar mass are observed in the literature \citep[e.g.][]{ChenWild2010b, ChenZ_DESIMgII_2025}, our results are likely insensitive to this due to the high stellar mass of the CMASS galaxy sample, significantly above the characteristic stellar mass of galaxies at these redshifts.

Within 1\,Mpc of the galaxy we find a very strong variation in absorber strength when comparing massive galaxies with different recent star formation histories (Fig.~\ref{fig:eqw_b_class}). Galaxies which have ongoing star formation (as evidenced by their stellar continuum shape) exhibit stronger \mgii absorption compared to quiescent galaxies. This is in broad agreement with previous results, although direct comparison is difficult due to the differing impact parameters, absorber strengths and galaxy stellar masses probed by different studies, and the different methods of separating galaxies into star-forming and quiescent \citep[e.g.][]{ChenWild2010b, Bordoloi2011, Lan2018, Anand2021, Lundgren2021, Wu2025, ChenZ_DESIMgII_2025}.

\subsection{\mgii absorption around post-starburst galaxies}
The first main result of this paper is that post-starburst galaxies, that have recently undergone a starburst followed by rapid quenching (see \citealt{French2021} for a review), show significantly stronger \mgii absorption strengths within a few hundred kpc of the galaxy compared to star-forming or quiescent galaxies (Fig.~\ref{fig:eqw_b_class}). Our results are complementary to those of \citet{Borthakur2013} and \citet{Heckman2017} who studied the halos around a small number of local starburst and post-starburst galaxies using multiple UV absorption lines, finding excess metal absorption and broader Ly$\alpha$ line widths.  While post-starburst galaxies are rare, this does not mean they are unimportant for understanding galaxy evolution in general: their transient spectral features are only visible for $\lesssim1$\,Gyr following the starburst \citep[e.g.][]{Wild2020}, and their number density is sufficient to account for 25-100\% of quenched galaxies, depending on mass and redshift \citep[e.g.][]{Wild2009,Wild2016,Rowlands2018,Belli2019,Zheng2022}. The CGM around rapidly quenched quiescent galaxies has implications for our understanding of the physical processes that cause the cessation of star formation in galaxies generally. 

There are several possibilities for the cause of the excess \mgii absorption around post-starburst galaxies: (1) gas is left over from the event that caused the starburst prior to the post-starburst phase,  (2) the environment of post-starburst galaxies is more crowded, with more nearby satellites and neighbours, (3) outflows drive gas out of the galaxies or impact the pre-existing CGM, (4) the state of the gas within the dark matter halo of post-starburst galaxies is more conducive to condensation of cool gas.  We address the first three of these in the following paragraphs, and leave the remaining possibility to Section \ref{sec:disc:simn}.

Analysis of deep imaging indicates that \emph{massive} post-starburst galaxies, as studied here, are caused predominantly by mergers \citep{Ellison2024}, with a high fraction showing tidal substructure \citep{Pawlik2018, Wilkinson2022, Verrico2023}. However, tidal substructure in massive galaxies typically extends to $\lesssim100$\,kpc even in the deepest imaging studies \citep{Duc2015}, including in 2 SDSS selected massive $z\sim0.7$ post-starburst galaxies \citep{DOnofrio2025}, i.e. the innermost data point probed in this paper, and therefore it seems unlikely that tidally stripped cold gas could be the primary origin for the excess absorption. Additionally, inspection of the individual spectra contributing to the innermost stacks presented in Fig. \ref{fig:classstacks} does not indicate that the absorption signal is driven by a small number of strong absorbers, as might be expected if some sightlines happened to pass through a tidal tail. Many (local) post-starburst galaxies still contain gas reservoirs, observed either in CO \citep{Rowlands2015,French2015} or \HI \citep{Zwaan2013,Ellison2025}. However, where spatially resolved observations are available, the gas is generally compact and associated with the central starburst \citep{Smercina2022,Otter2022}, and therefore can not be directly linked to an extended CGM.  

Despite their initial discovery in clusters \citep{DresslerGunn1983}, post-starburst galaxies are predominantly found in the field \citep{Quintero2004, Pawlik2018}. The  clustering of high mass post-starburst galaxies is consistent with star-forming galaxies of the same stellar mass, as expected given that star-forming galaxies are their likely progenitors at any given redshift \citep{Wilkinson2021}. Although deep imaging studies to measure low mass satellite galaxy numbers around high mass post-starburst galaxies have not been carried out to our knowledge, stacked imaging reveals that high mass post-starburst galaxies are highly compact with a deficit of light at large radii \citep{Maltby2018}. This deficit of light is contrary to what might be expected if there were an excess of dwarf satellites around post-starburst galaxies, and similarly argues against significant tidal debris. Thus, it seems unlikely that an excess of satellite galaxies is responsible for the excess \mgii\ absorption around post-starburst galaxies. 

Post-starburst galaxies are known to have cool gas outflows with high velocities of up to $\sim$1000\,km\,s$^{-1}$, detected via blueshifted \mgii\ or NaD absorption along the line-of-sight to the galaxy \citep{Tremonti2007,Coil2011,Maltby2019,Sun2024,Taylor2024}. Spectral fitting of the stellar continuum provides a robust age for the recent starburst of $\lesssim$1\,Gyr \citep[e.g.][]{Wild2020}, which provides (just enough) time for the outflow to travel $\sim$1\,Mpc. This is consistent with our results, where a  significant excess of \mgii\ absorption is detected out to several hundred kpc from post-starburst galaxies, compared to star-forming or quiescent galaxies, but only if the outflow does not slow or dissipate as it propagates. We turn to this question below.  Despite the low surface brightness of emission from extended galactic winds, integral field observations have recently allowed measurementts of extended ionised gas nebulae to $\sim$10's of kpc, outflowing at 100's of km/s \citep[e.g.][]{Burchett2021,Zabl2021}.  In one exteme case, extended ionised gas emission was detected to 100\,kpc from a starburst event that occured 0.4\,Gyr ago, likely a similar age to the post-starburst galaxies in this paper \citep{Rupke2019}. The combined observation of galactic winds around post-starburst galaxies in both absorption and emission has great potential that is as yet unexplored.

The possibility of a link between the extended cool CGM gas around post-starburst galaxies observed in this paper, and fast cool gas outflows observed previously in post-starburst galaxies is intriguing. It is of particular interest because the origin of fast, cool gas galactic outflows is still much debated and has significant implications for our understanding of supernovae and active galactic nuclei (AGN) feedback, as well as metal distribution in the CGM. It is neither clear what drives the winds, nor how far and fast they can propagate into the CGM before dissipating \citep[see][for a more complete discussion]{DiamondStanic2021,ThompsonHeckman2024}. An important factor for understanding the origin of the fast outflows in massive post-starburst galaxies is that these galaxies are very compact compared to quiescent galaxies of the same mass \citep[e.g.][]{Whitaker2012,Almaini2017,PoFeng2020,Setton2022,Zhang2024}, indicating a very compact preceding starburst, which is consistent with their merger origin. This initial compact starburst origin is also consistent with the highly compact, much younger ($\sim$10\,Myr) starburst or post-burst galaxies with very fast cool gas winds \citep{Tremonti2007, Sell2014, Whalen2022}, and the starburst regions in local LIRGS and ULIRGS \citep{Condon1991}. Analytic arguments and high resolution simulations suggest that stellar winds from very compact starburst regions can produce outflows with cool gas velocities of $\sim1000$\,km\,s$^{-1}$ \citep[e.g.][]{Heckman2011, Muratov2015, Heckman2017, Schneider2020}, although only out to $\sim10-100$\,kpc from the galaxy. An excess of \mgii\ absorption out to several hundred kpc from the galaxies may require additional mechanisms, for example, the acceleration of pre-existing CGM clouds via ram pressure from the hot wind fluid, dust destruction releasing metals into the CGM, or condensation from the hot gas due to thermal instabilities caused by the wind \citep{Heckman2017}.

On the other hand, the majority of cosmological simulations make simple sub-grid assumptions for star-formation driven wind velocities based on scaling relations for normal star-forming galaxies, leading to slower star-formation driven outflows than observed in compact starburst galaxies \citep[e.g.][]{Dave2019,Nelson2019}. For these simulations, AGN jets are used to drive the fastest outflows and enrich the distant CGM with metals \citep[e.g.][]{Yang2024}. Detailed merger simulations suggest that AGN driven jets triggered by gas rich mergers may drive material far out into the CGM around galaxies, potentially detectable as fast outflows \citep[e.g.][]{Talbot2024}. Number and energy density considerations also point to radio jets linking galaxy mergers to rapid quenching and the build up of massive queiscent galaxies throughout cosmic history \citep[e.g.][]{Heckman2024}. While a link between post-starburst galaxies and weak AGN is well established \citep[e.g.][]{Yan2006,Wild2007,Wild2010,Yesuf2014,Pawlik2018}, the post-starburst galaxies studied in this paper do not currently contain powerful optical AGN, due to the exclusion of bright QSO-like spectra from the sample selection \citep{Wild2010,Sell2014,Almaini2025}. A very small fraction may contain radio or X-ray bright AGN \citep[][Patil et al. submitted]{Almaini2025}.  In order for AGN to be the drivers of these outflows and the cause of the extended CGM, the AGN would need to ``flicker'' on and off for long after the starburst declines, perhaps due to instabilities driving gas into the central regions following the coalescence of the merging galaxies, or by winds from low mass stars \citep{Wild2010,Hopkins2012}. \citet{Almaini2025} argued that the number of X-ray bright AGN in post-starburst galaxies were sufficient to drive the fast outflows, given a sufficiently short duty cycle. \citet{Krishna2025} also finds that powerful optical quasars are 28 times more likely to be hosted by post-starburst galaxies than a control sample of mass-matched galaxies, further strengthening the link between post-starburst galaxies and powerful AGN. However, the presence of a powerful AGN is generally found to ionise the gas in simulations \citep[][see below]{Hani2018,Appleby2021}, and observational clustering studies indicate that quasars destroy \mgii\ and \HI\ absorbers to comoving distances of $\sim$1\,Mpc along their lines of sight \citep{Scott2000,Wild2008}, although this effect is likely anisotropic \citep{Farina2013,Lau2018}. Thus, there remains no direct observational evidence that AGN are driving the outflows and some arguments against AGN as the driving mechanism. 

Clearly, further observational and theoretical work is warranted to understand the link between the high velocity down-the-barrel outflows and the extended cool CGM gas properties of post-starburst galaxies, and whether AGN are indeed a requirement for their existence. For the moment, we conclude that it seems highly likely the extended \mgii haloes around post-starburst galaxies are related to the observed $\sim1000$\,km/s outflows. However, further theoretical work is required to determine both the driving mechanism and the means by which the clouds persist or are created out to hundreds of kpc from the galaxies.

\subsection{\mgii absorption beyond the halo virial radius}

\begin{figure}
    \centering
    \includegraphics[width=\columnwidth]{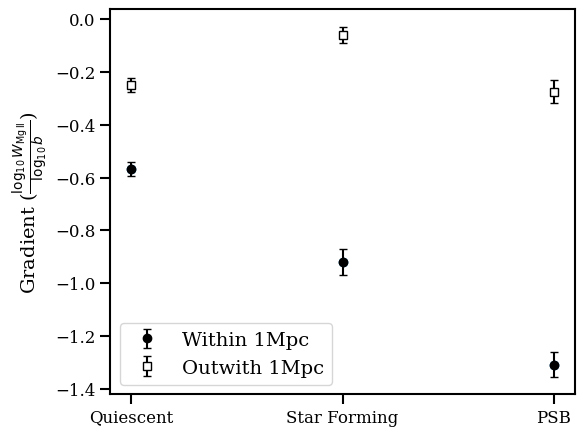}
    \caption{The gradient of \mgii absorption equivalent width (\AA) vs. impact parameter (kpc) within 1\,Mpc (circle) and outwith 1\,Mpc (square), as determined from the linear fits shown in Fig.~\ref{fig:eqw_b_class}, as a function of galaxy spectral type. Within 1\,Mpc, the post-starburst galaxies have the most rapid decline in \mgii absorption with distance from the galaxy, followed by star-forming and then quiescent. Outwith 1\,Mpc the decline is much shallower, perhaps indicating the transition between CGM and IGM, with the star-forming galaxies showing a nearly flat gradient at these distances. }
    \label{fig:gradients}
\end{figure}

The second main result of this paper is the clear flattening of the decline in equivalent width of \mgii beyond 1\,Mpc in all samples, and the fact that equivalent width becomes relatively independent of galaxy star formation history at this distance. In Figure \ref{fig:gradients} we compare the gradients of the best-fit lines for \mgii strength vs. impact parameter seen in Fig.~\ref{fig:eqw_b_class} within and outwith 1\,Mpc, highlighting a clear distinction between these two regions in all subsamples, although most notably in the star-forming and post-starburst galaxy samples. Beyond a projected distance of $\sim$1\,Mpc, the strength  of \mgii absorption becomes independent of galaxy type and the gradient flattens, which may indicate the expected transition from CGM to IGM. This transition occurs close to the expected virial radius of the galaxy haloes: central galaxies with stellar masses $\log_{10}(\mgal/\modot)\sim11-11.5$ are expected to reside in dark matter haloes with masses $\log_{10}({\rm M_h/M_\odot})\sim12.5-13.7$ at the redshift of our sample \citep{Shan2017}, although this varies significantly between different measured stellar mass to halo mass relations\footnote{\citet{Shan2017} studied CMASS galaxies, using the Portsmouth stellar mass value added catalogue. These stellar masses are 0.5\,dex smaller than the Granada value added stellar masses presented in Fig.~\ref{fig:galaxy_properties}, thus this is the appropriate mass range for our sample}. This corresponds to a virial radius of $\sim0.4-1.2$\,Mpc, which is consistent with the 1\,Mpc estimated from our data. Unfortunately, due to the sparsity of the CMASS galaxy observations, it is not possible to separate central and satellite galaxies and therefore take this analysis further. 

The transition is not apparent in our full dataset (Fig.~\ref{fig:eqw_b}), which is dominated by the quiescent galaxies where the effect is the smallest. This is consistent with previous results which focus on massive, predominantly quiescent, galaxies \citep{Zhu2014,PerezRafols2015}. \citet{Zhu2014} note a similar slight flattening in \mgii\ absorption strengths at 1\,Mpc around massive luminous red galaxies, which is fit well by the transition between 1 halo and 2 halo terms in the galaxy dark matter halo model. \citet{Kauffmann2017} detect excess \mgii\ out to at least 20\,Mpc, i.e. well beyond the galaxy haloes, pointing out that the extended profile of cool gas absorbers provides strong constraints on plausible feedback mechanisms. \citet{ChenZ_DESIMgII_2025} observe a flattening in absorption strength at $\sim$100-200\,kpc, consistent with the virial radius for their lower mass star-forming galaxies. Previous results comparing \mgii\ absorption around star-forming and quiescent galaxies have similarly found a difference only out to $\sim$100\,kpc \citep[e.g.][]{Lan2014,Lan2018}, which may again reflect the lower stellar masses of the emission line galaxies studied. Further work investigating the extent of cool absorber haloes around galaxies, as a function of stellar mass, central vs. satellite membership, and star formation histories could be attempted with larger upcoming surveys such as DESI.

\subsection{Comparison with simulations}\label{sec:disc:simn}

Cosmological simulations agree qualitatively on the decreasing presence of \mgii\ with distance from the galaxy, as characterised either via e.g. column density or covering fraction \citep{Huang2019, Ho2020, Nelson2020, Cook2024} or via synthetic absorption spectra \citep[e.g.][]{Ford2016,Liang2018,Appleby2021}, with the decrease corresponding to a decreasing number density of cool gas clouds. However, the extent and strength of absorption depends sensitively on assumed simulation parameters such as ionising background \citep{Liang2018,Hani2018,Huang2019,Appleby2021}, resolution \citep{Liang2020,Nelson2020}, presence of magnetic fields \citep{Vandevoort2021}, viscosity and precise numerical formulation \citep{Huang2019}, and low ionisation absorbers are typically found out to significantly smaller distances than observed in this paper. More generally, while it is agreed that galactic processes can impact the CGM and IGM on Mpc scales, and therefore subsequent galaxy evolution, precisely how that happens is highly simulation dependent, suggesting that Mpc scale CGM/IGM properties could provide excellent constraints on feedback mechanisms, should the other more mundane numerical differences be brought under control \citep[e.g.][]{Suresh2015,Yang2024}.

Cosmological simulations also find an increased \mgii\ absorption around star-forming compared to quiescent galaxies \citep{Ho2020, Appleby2021, Appleby2023}, with the difference corresponding to the increased cool gas content in the CGM around star-forming galaxies compared to quiescent galaxies of the same stellar mass. In the models of \citet{Appleby2021} this increased cool gas content appears to be due to ongoing star-formation feedback around star-forming galaxies enriching the CGM, alongside a reduction in jet-mode AGN feedback \citep[see also][]{Yang2024}. \citet{Appleby2023} find that green-valley galaxies, of which post-starburst galaxies are a small fraction \citep{Rowlands2018}, have similar CGM properties to quenched galaxies. The authors suggest that the quenching of the CGM and cessation of star formation occur together. This is not consistent with our observations of massive post-starburst galaxies where quenching has already occurred while the cool CGM gas is present in excess. Recent mergers are seen to increase the extent and metallicity of the CGM gas in simulations via outflows, although the concurrent presence of a powerful AGN causes this gas to be ionised well beyond the ionisation level of \mgii \citep{Hani2018}. Analysis of the CGM around specifically post-starburst galaxies in cosmological simulations would be an obvious next step, although potentially limited by resolution and sub-grid recipes.  

Given that the exact physical processes responsible for the cool gas phase in the CGM of even normal galaxies are not well understood \citep{Tumlinson2017, FaucherGiguere2023}, it is difficult to use results from simulations to reach any firm conclusions about the nature of the excess \mgii\ absorption around massive post-starburst galaxies, nor how it is linked to the fast outflow velocities seen along the line-of-sight from the galaxies. For example, if related to the outflows, is it cool gas that has been pushed out of the ISM (entrained), or has it condensed out of the hot and turbulent outflowing gas? Or is this extended CGM gas formed in the halo and unrelated to the outflows? Simulations of normal galaxies find that the cool gas in halos that is responsible for \mgii\ absorption condenses from the hotter halo gas due to thermal instabilities caused by strong density perturbations \citep{Nelson2020,Appleby2023}, rather than being due to e.g. infalling satellites \citep{Roy2024} or direct expulsion of cool gas clouds from the ISM. However, the post-starburst galaxies studied here may have very different physical processes impacting their CGM and, as discussed above, the physics of outflows is as poorly understood as the physics of the CGM. 

The final option is that the extended \mgii\ absorption is unrelated to the observed outflows, and instead related to the likelihood that the galaxies have experienced a recent major, gas-rich merger. We might speculate that the dynamical impact of a recent major merger on the halo could lead to increased instabilities and therefore increased condensation of cool CGM material around post-starburst galaxies \citep{Choudhury2019}. However, this is highly speculative, and detailed simulations would be required to demonstrate the plausibility of this scenario. Clearly there is still much work to be done on the theoretical side to understand the baryon cycle of galaxies, and the interplay between cool CGM gas and recent star formation histories of galaxies provides the potential for new constraints on feedback mechanisms. 

\section{Summary}\label{sec:summary}

We measured the strength of \mgii\ absorption using stacked sightlines to background quasars from the SDSS DR7 quasar catalogue, passing close to galaxies with masses $\log_{10}({\rm M/M_\odot})\gtrsim11$, with different recent star formation histories from the SDSS DR12 CMASS catalogue. This enabled us to probe the cool gas in the CGM of galaxies out to 9\,Mpc. The main findings of this work are:

\begin{enumerate}
\item We recovered the expected decrease in \mgii\ equivalent width with increasing distance from the galaxies (Figs. \ref{fig:allstacks} and \ref{fig:eqw_b}), as well as the expected enhancement in \mgii\ in star-forming compared to quiescent galaxies (Figs. \ref{fig:classstacks} and \ref{fig:eqw_b_class}).
\item We found significantly stronger \mgii absorption within several hundred kpc of post-starburst galaxies compared to quiescent or star-forming galaxies (Figs. \ref{fig:classstacks} and \ref{fig:eqw_b_class}). This extended cool CGM gas may be related to the previously observed high velocity ($\sim$1000\,km/s) outflows along the line-of-sight to post-starburst galaxies. This result has important implications for our understanding of why galaxies stop forming stars, as well as how metals are dispersed into the CGM. 
 \item Beyond 1\,Mpc we observe the decrease in equivalent width to flatten and become largely independent of galaxy type (Figs. \ref{fig:eqw_b_class} and \ref{fig:gradients}). This distance is close to the expected location of the virial radius of the galaxies, and thus the transition between CGM and IGM.
\end{enumerate}

Because post-starburst galaxies represent a significant pathway for the formation of quenched, elliptical galaxies our results have important implications for understanding galaxy evolution more generally. This study demonstrates that the recent past star formation history of galaxies significantly correlates with the properties of their CGM, and emphasise the importance of the precise form of ongoing and past star formation activity, i.e. continuous vs. bursty, rather than a simple measure of ongoing star formation rate. Understanding the cause of the observed correlations will require further theoretical work, particularly with respect to the generation of cool, low ionisation, extended gas haloes in simulated post-starburst galaxies. While we speculate that the extended CGM haloes are related to the fast outflows observed in massive post-starburst galaxies, and there are several plausible mechanisms for generating the fast winds, there is no consensus on how they would lead to excess cool gas in halos at such large distances from the galaxies. Future large spectroscopic surveys will enable the expansion of the dataset to include more post-starburst galaxies, which could then be split into further subsamples such as stellar mass, starburst age and environment. This will help refine our understanding of the interplay between galaxy evolution, the CGM, and the IGM, as well as their roles in shaping \mgii absorption across cosmic scales.

\section*{Acknowledgements}

We would like to thank the anonymous referee for the thorough review which improved the manuscript, as well as input from the community following submission to the arxiv. VW acknowledges Science and Technologies Facilities Council (STFC) grant ST/Y00275X/1 and Leverhulme Fellowship RF-2024-589/4. 

Funding for the Sloan Digital Sky Survey (SDSS) and SDSS-II has been provided by the Alfred P. Sloan Foundation, the Participating Institutions, the National Science Foundation, the U.S. Department of Energy, the National Aeronautics and Space Administration, the Japanese Monbukagakusho, and the Max Planck Society, and the Higher Education Funding Council for England. The SDSS Web site is http://www.sdss.org/.

The SDSS is managed by the Astrophysical Research Consortium (ARC) for the Participating Institutions. The Participating Institutions are the American Museum of Natural History, Astrophysical Institute Potsdam, University of Basel, University of Cambridge, Case Western Reserve University, The University of Chicago, Drexel University, Fermilab, the Institute for Advanced Study, the Japan Participation Group, The Johns Hopkins University, the Joint Institute for Nuclear Astrophysics, the Kavli Institute for Particle Astrophysics and Cosmology, the Korean Scientist Group, the Chinese Academy of Sciences (LAMOST), Los Alamos National Laboratory, the Max-Planck-Institute for Astronomy (MPIA), the Max-Planck-Institute for Astrophysics (MPA), New Mexico State University, Ohio State University, University of Pittsburgh, University of Portsmouth, Princeton University, the United States Naval Observatory, and the University of Washington.

Funding for SDSS-III has been provided by the Alfred P. Sloan Foundation, the Participating Institutions, the National Science Foundation, and the U.S. Department of Energy Office of Science. The SDSS-III web site is http://www.sdss3.org/. SDSS-III is managed by the Astrophysical Research Consortium for the Participating Institutions of the SDSS-III Collaboration including the University of Arizona, the Brazilian Participation Group, Brookhaven National Laboratory, Carnegie Mellon University, University of Florida, the French Participation Group, the German Participation Group, Harvard University, the Instituto de Astrofisica de Canarias, the Michigan State/Notre Dame/JINA Participation Group, Johns Hopkins University, Lawrence Berkeley National Laboratory, Max Planck Institute for Astrophysics, Max Planck Institute for Extraterrestrial Physics, New Mexico State University, New York University, Ohio State University, Pennsylvania State University, University of Portsmouth, Princeton University, the Spanish Participation Group, University of Tokyo, University of Utah, Vanderbilt University, University of Virginia, University of Washington, and Yale University.

%%%%%%%%%%%%%%%%%%%%%%%%%%%%%%%%%%%%%%%%%%%%%%%%%%
\section*{Data Availability}

All data used in this paper are publicly available. SDSS data and analysis products can be downloaded from the SDSS Science Archive server\footnote{\url{https://data.sdss.org/sas/}}.

%%%%%%%%%%%%%%%%%%%% REFERENCES %%%%%%%%%%%%%%%%%%
\bibliography{SDSSPaper}

%%%%%%%%%%%%%%%%%%%%%%%%%%%%%%%%%%%%%%%%%%%%%%%%%%

%%%%%%%%%%%%%%%%% APPENDICES %%%%%%%%%%%%%%%%%%%%%

\appendix

\renewcommand\thefigure{\thesection.\arabic{figure}}

\section{Galaxy stacked spectra}\label{app:stacks}
\setcounter{figure}{0}
\begin{figure*}
   \centering
    \includegraphics[width=\textwidth]{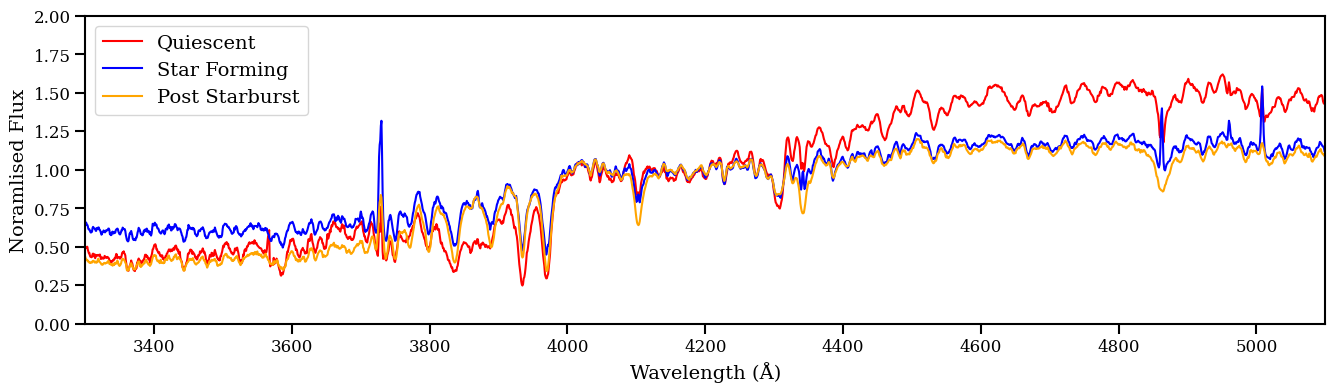}
    \caption{The stacked spectra of galaxies in the different mean-field ICA classes: quiescent (red), star-forming (blue) and post-starburst (orange). Prior to stacking all spectra were normalised to a mean flux of unity between 4150 and 4250\AA. }
    \label{fig:stacks}
\end{figure*}

Fig.~\ref{fig:stacks} shows the stacked galaxy spectra in each of the MFICA classes: post-starburst, star-forming, and quiescent galaxies within the spectral region that the MFICA uses to classify the spectra. This clearly highlights the features and shapes that the MFICA method is using to identify different types of galaxies. The quiescent galaxies have a strong 4000\AA\ break and red spectrum, the star-forming galaxies have emission lines and a blue spectrum, while the post-starburst galaxies have a Balmer break, strong Balmer absorption lines, very weak or no emission lines and a significantly different continuum shape compared to either star-forming or quiescent galaxies.

\section{The impact of redshift and stellar mass on equivalent width vs. impact parameter}\label{app:systematics}

\setcounter{figure}{0}
\begin{figure*}
    \centering
    \includegraphics[width=\textwidth]{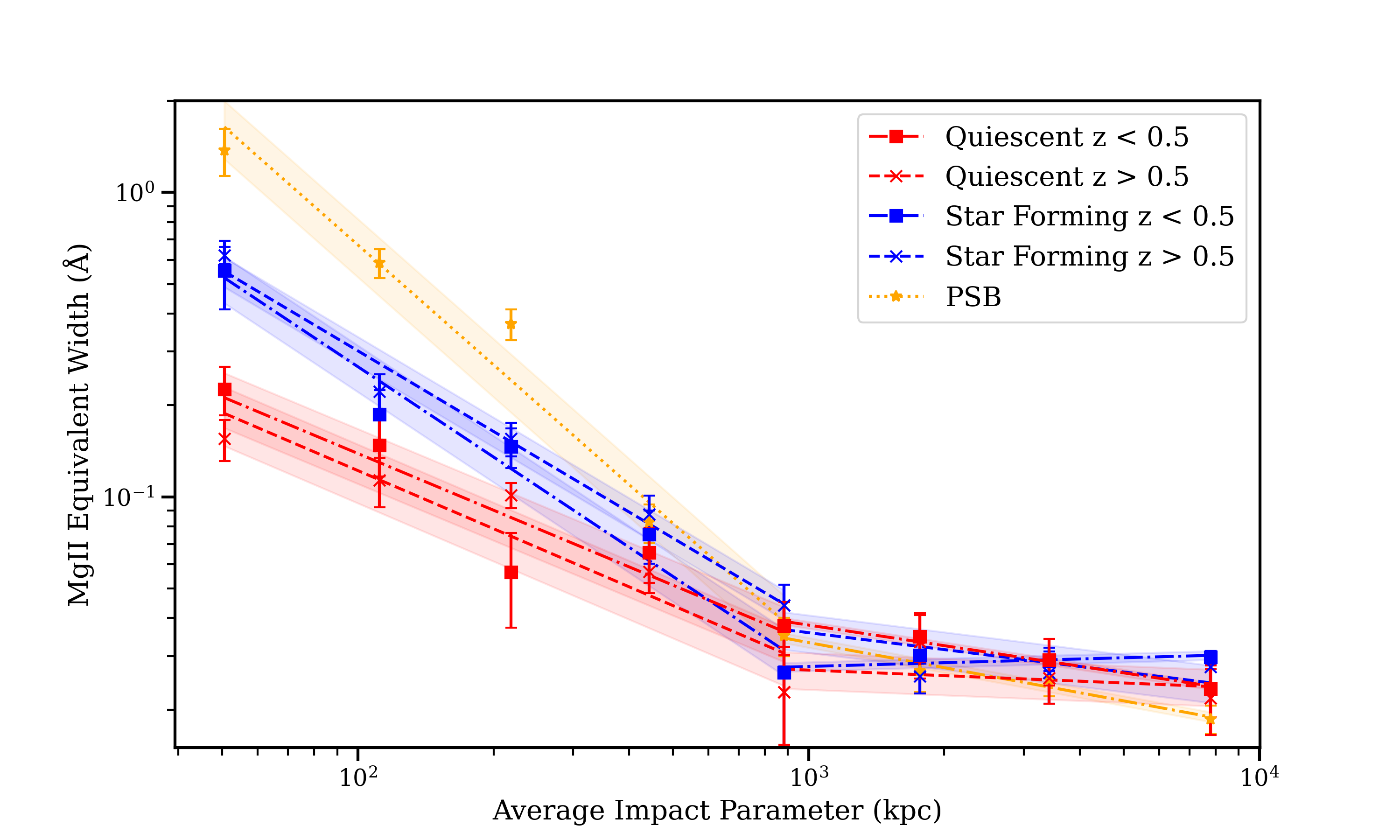}
    \caption{The equivalent width of the \mgii absorption doublet as a function of average impact parameter for galaxies of different spectral types: post-starburst (orange), star-forming (blue) and quiescent (red). The star-forming and quiescent subsamples are split by redshift, with dashed lines for $z>0.5$ and dot-dash lines for $z<0.5$. The post-starburst sample is the same as in Fig.~\ref{fig:eqw_b_class}.}
    \label{fig:eqw_b_class_redshift}
\end{figure*}

\begin{figure*}
    \centering
    \includegraphics[width=\textwidth]{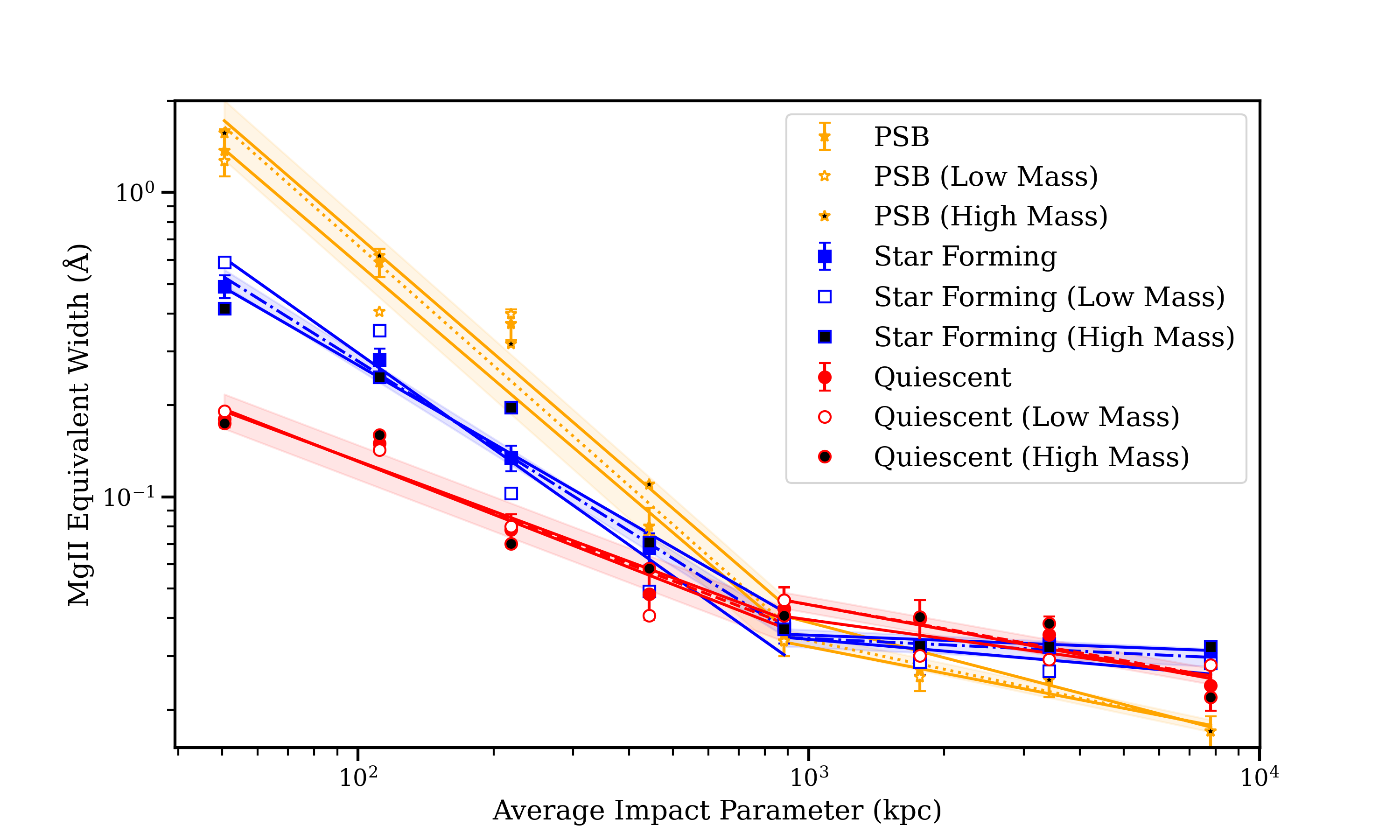}
    \caption{The equivalent width of the \mgii absorption doublet as a function of average impact parameter for galaxies of different spectral types: post-starburst (orange), star-forming (blue) and quiescent (red). All subsamples are split above and below the median stellar mass in each sub-sample, with solid lines for the low and high mass samples, and dotted/dashed lines for the full samples.}
    \label{fig:eqw_b_class_mass}
\end{figure*}

In this section, we first check for potential redshift evolutionary effects that may dilute or strengthen the relation between the equivalent width of \mgii and stellar population at fixed impact parameter, because of the differences in redshift distributions between the three samples (Fig.~\ref{fig:redshifts}). In Fig.~\ref{fig:eqw_b_class_redshift} we  repeat our analysis, splitting the star-forming and quiescent samples by redshift, to demonstrate the small changes in absorption strength over the redshift range studied here ($0.35<z<0.8$) for these very massive galaxies. There is no strong trend between the equivalent width of \mgii  and redshift, and the observed gradients do not significantly alter. We also calculated the stacks by (a)  weighting the input spectra of the star-forming and quiescent subsamples to match the redshift distribution of the post-starburst galaxies, and (b) sub-sampling the star-forming and quiescent samples to have redshift distributions to match that of the post-starburst galaxies. In both cases, there was no qualitative change in our results. The sub-sampled version provides an additional check on the bootstrap errors in the bins with smaller numbers of objects, as we are able to sample from a larger pool of spectra than is used to create each individual stack. As expected, we find the errors to increase slightly, but this does not impact the overall result. Regardless of method,  absorption strength correlates strongly with the recent star formation history of galaxies at fixed impact parameter.

To investigate whether we might expect our results to be qualitatitively different if we were able to normalise by a stellar mass scaled quantity, such as the halo virial radius, in Fig.~\ref{fig:eqw_b_class_mass} we show that there is no significant difference between the equivalent width vs. impact parameter when samples are split at their median stellar mass. We additionally investigated trends with mass alone for the full sample, finding that mass does not determine absorption strength in this high mass galaxy sample.

%%%%%%%%%%%%%%%%%%%%%%%%%%%%%%%%%%%%%%%%%%%%%%%%%%

\end{document}